\documentclass[runningheads]{eptcs}
\usepackage[utf8]{inputenc}
\usepackage[english]{babel}
\usepackage[T1]{fontenc}
\usepackage{graphicx}
\usepackage{listings}
\usepackage{xcolor}
\usepackage{pgf,tikz}
\usepackage{wrapfig}

\usepackage[counterclockwise]{rotating}
\usepackage[english]{varioref}
\usepackage{url}
\usepackage{xypic}
\xyoption{all}
\usepackage{varioref}

\usetikzlibrary{knots,calc,shapes.geometric}
\usetikzlibrary{arrows}

\definecolor{dkgreen}{rgb}{0,0.6,0}
\definecolor{ltblue}{rgb}{0,0.4,0.4}
\definecolor{dkblue}{rgb}{0,0.4,0.4}
\definecolor{dkviolet}{rgb}{0.3,0,0.5}
\definecolor{dkred}{rgb}{1.0,0.0,0.0} 

\definecolor{ttzzqq}{rgb}{0.2,0.6,0.0} 
\definecolor{ffqqqq}{rgb}{1.0,0.0,0.0} 
\definecolor{qqqqff}{rgb}{0.0,0.0,1.0} 
\definecolor{ffxfqq}{rgb}{1.0,0.4980392156862745,0.0} 
\definecolor{xfqqff}{rgb}{0.4980392156862745,0.0,1.0} 
\definecolor{zzttqq}{rgb}{0.6,0.2,0.0} 
\definecolor{ttqqqq}{rgb}{0.2,0.0,0.0}
\definecolor{mygreen}{RGB}{34,120,15}
\newcommand{\rk}[1]{rk\{{#1}\}}
\lstdefinelanguage{Coq}{ 
    mathescape=true,
    texcl=false, 
    morekeywords=[1]{Section, Module, End, Require, Import, Export,
        Variable, Variables, Parameter, Parameters, Axiom, Hypothesis,
        Hypotheses, Notation, Local, Tactic, Reserved, Scope, Open, Close,
        Bind, Delimit, Definition, Let, Ltac, Fixpoint, CoFixpoint, Add,
        Morphism, Relation, Implicit, Arguments, Unset, Contextual,
        Strict, Prenex, Implicits, Inductive, CoInductive, Record,
        Structure, Canonical, Coercion, Context, Class, Global, Instance,
        Program, Infix, Theorem, Lemma, Corollary, Proposition, Fact,
        Remark, Example, Proof, Goal, Save, Qed, Defined, Hint, Resolve,
        Rewrite, View, Search, Show, Print, Printing, All, Eval, Check,
        Projections, inside, outside, Def},
    morekeywords=[2]{forall, exists, exists2, fun, fix, cofix, struct,
        match, with, end, as, in, return, let, if, is, then, else, for, of,
        nosimpl, when},
    morekeywords=[3]{Type, Prop, Set, true, false, option},
    morekeywords=[4]{pose, set, move, case, elim, apply, clear, hnf,
        intro, intros, generalize, rename, pattern, after, destruct,
        induction, using, refine, inversion, injection, rewrite, congr,
        unlock, compute, ring, field, fourier, replace, fold, unfold,
        change, cutrewrite, simpl, have, suff, wlog, suffices, without,
        loss, nat_norm, cut, trivial, revert, bool_congr, nat_congr,
        symmetry, transitivity, auto, split, left, right,
        autorewrite, clear_all_rk, my_inO,
      intuition, solve_hyps_min,  solve_hyps_max, constructor, assumption},
    morekeywords=[5]{by, done, exact, reflexivity, tauto, romega, omega,
        assert, solve, contradiction, discriminate},
    morekeywords=[6]{do, last, first, try, idtac, repeat},
    morecomment=[s]{(*}{*)},
    showstringspaces=false,
    morestring=[b]",
    morestring=[d]’,
    tabsize=3,
    extendedchars=false,
    sensitive=true,
    breaklines=false,
    basicstyle=\small,
    captionpos=b,
    columns=[l]flexible,
    identifierstyle={\ttfamily\color{black}},
    keywordstyle=[1]{\ttfamily\color{dkviolet}},
    keywordstyle=[2]{\ttfamily\color{dkgreen}},
    keywordstyle=[3]{\ttfamily\color{ltblue}},
    keywordstyle=[4]{\ttfamily\color{dkblue}},
    keywordstyle=[5]{\ttfamily\color{dkred}},
    stringstyle=\ttfamily,
    commentstyle={\ttfamily\color{dkgreen}},
    literate=
    {\\forall}{{\color{dkgreen}{$\forall\;$}}}1
    {\\exists}{{$\exists\;$}}1
    {<-}{{$\leftarrow\;$}}1
    {=>}{{$\Rightarrow\;$}}1
    {==}{{\code{==}\;}}1
    {==>}{{\code{==>}\;}}1
    {->}{{$\rightarrow\;$}}1
    {<->}{{$\leftrightarrow\;$}}1
    {<==}{{$\leq\;$}}1
    {\#}{{$^\star$}}1 
    {\\o}{{$\circ\;$}}1 
    {\@}{{$\cdot$}}1 
    {\/\\}{{$\wedge\;$}}1
    {\\\/}{{$\vee\;$}}1
    {~}{{$\sim$}}1
    {\@\@}{{$@$}}1
    {\\mapsto}{{$\mapsto\;$}}1
    {\\hline}{{\rule{\linewidth}{0.5pt}}}1
}[keywords,comments,strings]

\title{Towards Automatic Transformations of Coq Proof Scripts}
\author{Nicolas Magaud
\institute{Lab. ICube UMR 7357 CNRS Universit\'e de
  Strasbourg, France\\
  \email{magaud@unistra.fr}}}

\def\titlerunning{Towards Automatic Transformations of Coq Proof Scripts}
\def\authorrunning{N. Magaud}
\begin{document}
\maketitle              
\begin{abstract}
Proof assistants like Coq are increasingly popular to help
mathematicians carry out proofs of the results they
conjecture. However, formal proofs remain highly technical and are
especially difficult to reuse. In this paper, we present a framework to carry out
\textsl{a posteriori} script transformations. These transformations
are meant to be applied as an automated post-processing step, once the
proof has been completed. As an example, we present a
transformation which takes an arbitrary large proof script and produces
an equivalent single-line proof script, which can be executed by Coq in
one single step. Other
applications, such as fully expanding a proof script (for debugging purposes), removing
all named hypotheses, etc. could be developed within this
framework. We apply our tool to various Coq proof scripts, including
some from the GeoCoq library.


\end{abstract}
\section{Motivations}
Proof assistants like Coq~\cite{BC04,coqmanual} are increasingly popular to help
mathematicians carry out proofs of the results they
conjecture. However, formal proofs remain highly technical and are
especially difficult to reuse.
Once the proof effort is done, the proof scripts are left as they are
and they often break when upgrading to a more recent version of the prover.  To reduce the burden
of maintaining the proof scripts of Coq, we propose a tool to
post-process the proof scripts to make them cleaner and easier to reuse.
The first transformation that we focused on consists in compacting a
several-step proof script into a single-step proof script. 
Even though our framework can be used to implement other proof script
transformations, this one is of special interest to us. 
Indeed, we recently designed a prover for projective incidence
geometry~\cite{DBLP:conf/issac/BraunMS21,DBLP:journals/corr/abs-2201-00539} which relies on the
concept of rank to carry out proofs of geometric theorems such as
Desargues or Dandelin-Gallucci automatically. This prover produces a
trace (a large Coq proof script containing several  statements and
their proofs). We hope to use the proof transformation tool to shape
up the automatically generated
proofs and make them easier to reuse and integrate in larger proof
repositories.

More generally, proof maintenance and reuse tools  have
been studied extensively by Talia Ringer et
al. \cite{DBLP:conf/cpp/RingerYLG18,DBLP:journals/ftpl/RingerPSGT19}. Contrary
to our approach, their tools aim at fixing the issues when they occur. In our setting,
we think it is better to try and improve the proof scripts so that
they are less likely to break, even after several years and numerous
updates of the components.  

\paragraph{Outline of the paper}
The paper is organized as follows. In Sect.~\ref{sec:example}, we
present a simple
example of a proof script transformation. In Sect.~\ref{sec:implementation}, we describe the implementation of our
tool as well as the future extensions we currently develop.   In
Sect. ~\ref{sec:conclusion}, we present some concluding remarks and the perspectives
of this work.  
\section{Transforming Large Proof Scripts into One-line Scripts}\label{sec:example}
The Coq tactic language \cite{DBLP:conf/lpar/Delahaye00} features tacticals to execute some
tactics in a sequence
\mbox{\textcolor{ltblue}{\texttt{tac1;tac2;tac3}}} or to try and
execute different tactics on the same
goal \mbox{\textcolor{ltblue}{\texttt{solve [ tac1 | tac2 | tac3 ]}}}. Moreover these tacticals
can be combined. E.g.~\mbox{\textcolor{ltblue}{\texttt{tac0 ; [tac1 | tac2 | tac3]}}} runs the first
tactic \textcolor{ltblue}{\texttt{tac0}} which should yield 3 subgoals. The first one is
solved using \textcolor{ltblue}{\texttt{tac1}}, the second one using \textcolor{ltblue}{\texttt{tac2}} and the
third one using \textcolor{ltblue}{\texttt{tac3}}.
Once a proof script is written (as several steps) by the user, we can use these
tacticals to build an equivalent proof script, which can be executed
in a single step.

Let us consider a simple example, proving the distributivity of
the connective or (\textcolor{ltblue}{$\vee$}) 
over the connective and
(\textcolor{ltblue}{$\wedge$}) 
as shown in the
statement of figure~\ref{example}. 

\begin{figure}
{\small \begin{lstlisting}[language=Coq]
Lemma foo : forall A B C : Prop, A \/ (B /\ C) -> (A\/B)/\(A\/C).
\end{lstlisting}}

  \begin{minipage}{0.45\linewidth}
  {\small
  \begin{lstlisting}[language=Coq]
Proof.
  intros; destruct H. 
  split.
  left; assumption. 
  left; assumption. 
  destruct H.
  split. 
  right; assumption. 
  right; assumption. 
  Qed.
\end{lstlisting}
}
\end{minipage}
\begin{minipage}{0.45\linewidth}
  {
\small
  \begin{lstlisting}[language=Coq]
Proof.
intros; destruct H;
  [ split;
    [ left; assumption
    | left; assumption ]
  | destruct H ;
    split;
    [ right; assumption
    | right; assumption ] ].
Qed.
\end{lstlisting}
}
\end{minipage}
\caption{\label{example}A user-written script (left) and the equivalent
  single-step script (right)}
\end{figure}

The left-hand side presents the proof script that one may expect from a
master student, factorizing some parts but still decomposing the
reasoning in several steps. On the right-hand side, we propose a one-line
script to carry out exactly the same proof.

In Coq, writing directly the right-hand side is almost impossible, whereas it
is fairly easy to generate it automatically from the left-hand side.
In the Coq standard library, several lemmas are proved using a single one-line
tactic. The main advantage is that it provides concise and structured
proofs but it has the drawback that, when something goes wrong, it is
hard to debug and fix it.

\section{Experiments, Limitations and Results}\label{sec:implementation}
\subsection{Implementation}
We choose to implement our tool in \texttt{OCaml}, using the serialisation mechanism  \texttt{serapi}~\cite{GallegoArias2016SerAPI}
developed by Emilio Gallego Arias for communication with the Coq
proof assistant.  
Our tool uses anonymous pipes to communicate with \texttt{serapi}, which itself
sends requests to Coq and retrieves the answers.  
Commands are kept as in the input file. Tactics are aggregated
using tacticals such as \textcolor{ltblue}{\texttt{;}}, \textcolor{ltblue}{\texttt{[}} and \textcolor{ltblue}{\texttt{]}}.
At each step of the proof, we compare the current number of subgoals to the
number of subgoals right before the execution of the current
tactic. If it is the same, we simply concatenate the tactics with a
\textcolor{ltblue}{\texttt{;}} between them. If the number of goals increases, we open a
square bracket \textcolor{ltblue}{\texttt{[}} and push into the stack the previous number of
goals. Each time a goal is solved, we check whether some goals remain
to be proved at this level. If yes, we add another \textcolor{ltblue}{\texttt{;}} and
then focus on the next subgoal. If there are no more subgoals at this
level, we pop the 0 from the top of the stack, thus closing the
current level with a \textcolor{ltblue}{\texttt{]}} and carry on with subgoals of the
previous level. 

The source code\footnote{\url{https://github.com/magaud/coq-lint}} as
well as some examples are freely available online.
It is developped using \texttt{Coq} 8.17.0 and the corresponding
\texttt{serapi} version 8.17.0+0.17.0. 

\subsection{Limitations}
So far, commands and tactics are told apart simply by assuming
commands start with a capital letter [A-Z] and tactics with a small
letter [a-z]. This convention is well-known in Coq, however in some developments (e.g.~GeoCoq), some ad-hoc
user tactics may start with a capital letter. Handling this properly
requires additional developments and is currently under way.  

To make the transformation easier, a first phase could be added to our proof script transformer to remove
all commands which lay among the proof steps
(e.g.~\textcolor{dkviolet}{\texttt{Check}}, \textcolor{dkviolet}{\texttt{Print}} or \textcolor{dkviolet}{\texttt{Locate}}) and make sure
all tactics names start with a small letter.

Finally, Coq proof scripts can be structured using bullets
(\texttt{+}, \texttt{-}, \texttt{*}) as well as curly brackets to
identify some subproofs. In addition, one can direct work on a goal
which is not the current one using the \texttt{2: \textcolor{ltblue}{tac}.} notation
which performs the tactic \texttt{tac} on the second goal of the
subgoals.  We still need to devise a way to deal properly with such
partially-structured proof script.  

\subsection{Successful Transformations}
In addition to our test suite examples, we consider more challenging
proof scripts. We successfully transformed a library file from the Standard
Library of Coq:  Cantor.v\footnote{\url{https://github.com/coq/coq/blob/master/theories/Arith/Cantor.v}}
from the \texttt{Arith} library as well as some large files from the \texttt{GeoCoq} library
\cite{boutry:hal-01178236,narboux:inria-00118812}
(e.g.~\texttt{orthocenter.v}\footnote{\url{https://github.com/GeoCoq/GeoCoq/blob/master/Highschool/orthocenter.v}}). 
As the tool gets more mature, we plan to transform more files, and we
shall especially focus on the \texttt{GeoCoq} library which features
several different proof styles and thus shall allow us to evaluate the
robustness of our tool.

\subsection{Refactoring Proof Scripts Automatically Generated by our Prover for Projective Incidence
  Geometry}
We recently developed a new
way~\cite{DBLP:conf/issac/BraunMS21,DBLP:journals/corr/abs-2201-00539},
based on ranks, to automatically prove statements in projective incidence
geometry. Our approach works well but produces proof scripts which are
very large and often feature several auxiliary lemmas.
Figure~\ref{fig2} presents a very simple example \texttt{LABC}, which is formally
proven by our tool but yields a fairly verbose proof script using one
intermediate lemma \texttt{LABCD} (see appendix \ref{proofscript} for details).  

\begin{figure}
\begin{minipage}{0.22\linewidth}
    \begin{center}
 \scalebox{1.75}{\begin{tikzpicture}[line cap=round,line join=round,>=triangle 45,x=1.0cm,y=1.0cm]
    \tikzstyle{every node}=[font=\tiny]
    \clip(-1.0,-1.0) rectangle (1.0,1.0);
    \fill[color=white,fill=white,fill opacity=0.1] (-0.6770714035101904,-0.4325980198719555) -- (0.34110966716268676,0.6
899787854448113) -- (0.7603606962632832,-0.70853868710
0658) -- cycle;
    \fill[color=white,fill=white,fill opacity=0.1] (-0.6770714035101904,-0.4325980198719555) -- (-0.12848984015185505,0.
17223048911171895) -- (0.7603606962632832,-0.708538687
100658) -- cycle;
    \draw (-0.6770714035101904,-0.4325980198719555)-- (0.34110966716268676,0.6899787854448113);
    \draw (0.34110966716268676,0.6899787854448113)-- (0.7603606962632832,-0.708538687100658);
    \draw (-0.12848984015185505,0.17223048911171895)-- (0.7603606962632832,-0.708538687100658);
    \draw [color=zzttqq] (-0.6770714035101904,-0.4325980198719555)-- (0.34110966716268676,0.6899787854448113);
    \draw [color=zzttqq] (0.34110966716268676,0.6899787854448113)-- (0.7603606962632832,-0.708538687100658);
    \draw [color=zzttqq] (0.7603606962632832,-0.708538687100658)-- (-0.6770714035101904,-0.4325980198719555);
    \draw [color=zzttqq] (-0.6770714035101904,-0.4325980198719555)-- (-0.12848984015185505,0.17223048911171895);
    \draw [color=zzttqq] (-0.12848984015185505,0.17223048911171895)-- (0.7603606962632832,-0.708538687100658);
    \draw [color=zzttqq] (0.7603606962632832,-0.708538687100658)-- (-0.6770714035101904,-0.4325980198719555);
    \draw (0.7603606962632832,-0.708538687100658)-- (-0.6770714035101904,-0.4325980198719555);
    \begin{scriptsize}
    \draw [fill=black] (-0.6770714035101904,-0.4325980198719555) circle (1pt);
    \draw[color=black] (-0.8554417739004563,-0.4024269894667369) node {$A$};
    \draw [fill=black] (0.34110966716268676,0.6899787854448113) circle (1pt);
    \draw[color=black] (0.33795740378599054,0.864211946506821) node {$D$};
    \draw [fill=black] (0.7603606962632832,-0.708538687100658) circle (1pt);
    \draw[color=black] (0.9018833300302452,-0.7083676566954394) node {$B$};
    \draw [fill=black] (-0.12848984015185505,0.17223048911171895) circle (1pt);
    \draw[color=black] (-0.28064341960192262,0.2321681329971913) node {$C$};
    \end{scriptsize}
\end{tikzpicture}}
\end{center}
\end{minipage}
\begin{minipage}{0.77\linewidth}
  \begin{itemize}
  \item {Informal statement} \\
    Assume that ABD is a triangle,\\
   Assume that C is a point on AD, such that C$\neq$ A and C$\neq$ D,\\
  Then ABC is a triangle. \\

    \item {Expressed using ranks}
  $$\begin{array}{l}
\forall A,B,C,D :\texttt{Point},\\
      \rk{A, D, B } = 3 \rightarrow 
      \rk{A,C,D} = 2 \rightarrow\\
    \rk{C,A} = 2 \rightarrow
    \rk{C,D} = 2 \rightarrow\\
    \rk{A,C,B} = 3.
    \end{array}
    $$
    \end{itemize}
  \end{minipage}
\caption{\label{fig2}An example of a statement in projective geometry, formalized
  using ranks}
\end{figure}
We plan to use our script transformation tool to refactor
automatically generated proof scripts, inlining auxiliary lemmas and thus making
proof scripts more concise and hopefully more readable for humans.  
\subsection{Next steps}
To fully evaluate the tool, we need to handle larger examples, outside
of the standard library of Coq. The next
step consists in improving Coq options handling (e.g.~\texttt{-R}) to
our script transformation tool to tackle other formal proof
libraries. 

We  also plan to propose the reciprocal script transformation, turning
a single-step proof script into a more detailed (easier to debug)
proof script. This could especially be useful when porting formal
proofs from one version of Coq to the next one.

Other applications of
interest could be to remove the names of all variables or hypotheses from the
scripts, or at least to force them to be explicitly introduced.
The script snippet \texttt{\textcolor{ltblue}{intros}; \textcolor{ltblue}{apply} H} could be replaced by a
more precise one \texttt{\textcolor{ltblue}{intros} n p H; \textcolor{ltblue}{apply} H}. This way, we could
ensure that the proofs are not broken when the names of automatic
variables change. From a reliability point of view, it would be even
better to use the tactic \texttt{\textcolor{ltblue}{intros}; \textcolor{ltblue}{assumption}}. Although its cost is
higher (because we need to search the correct hypothesis among all of
them every time we run the tactic), it does not depend on some arbitrary variable names.  

Finally, regarding our current implementation, it would be
interesting to benchmark the transformation to see whether
transforming the whole standard library of Coq into single-step proof
scripts could improve the compilation time of this library.

\section{Conclusions and perspectives}\label{sec:conclusion}
We build a proof script transformation tool, which transforms an
arbitrary large proof script into a single-step
\textsl{one-Coq-tactic} proof script. This tool has been successfully
experimented on some significant library files from the Coq ecosystem.

This first example shows that the approach is sound and we plan to
extend it to integrate tactic languages such as
ssreflect~\cite{DBLP:journals/jfrea/GonthierM10}, Ltac2~\cite{Pedrot2019} or Mtac~\cite{DBLP:journals/jfp/ZilianiDKNV15} in our framework.  In the longer term, we
expect to design some new proof script transformations and combine
them in order to build more reliable proof developments which can
last longer and would be easier to maintain. Among these
transformations, we shall start with a mechanism to transform a proof
script into a sequence of atomic proof steps (to make debugging easier
when the proof breaks). We may also study how to transform
proofs carried out automatically by their actual traces, avoiding
recomputing the proof search each time the proof is re-runned.

%
%
%
\bibliographystyle{splncs04}
%

\label{sect:bib}
\bibliography{references.bib}

\appendix
\section{\label{proofscript}Proof Script for our basic example}
\begin{lstlisting}[language=Coq]
Lemma LABCD : forall A B C D ,
rk(A:: C::nil) = 2 -> rk(A:: B:: D::nil) = 3 -> 
rk(C:: D::nil) = 2 -> rk(A:: C:: D::nil) = 2 -> 
rk(A:: B:: C:: D::nil) = 3.
Proof.
intros A B C D 
HACeq HABDeq HCDeq HACDeq .
assert(HABCDm2 : rk(A:: B:: C:: D:: nil) >= 2).
{
	assert(HACmtmp : rk(A:: C:: nil) >= 2) 
	       by (solve_hyps_min HACeq HACm2).
	assert(Hcomp : 2 <= 2) by (repeat constructor).
	assert(Hincl : incl (A:: C:: nil) (A:: B:: C:: D:: nil)) 
	       by (repeat clear_all_rk;my_inO).
	apply (rule_5 (A:: C:: nil) (A:: B:: C:: D:: nil) 2 2 HACmtmp Hcomp Hincl).
}
assert(HABCDm3 : rk(A:: B:: C:: D:: nil) >= 3).
{
	assert(HABDmtmp : rk(A:: B:: D:: nil) >= 3) 
	       by (solve_hyps_min HABDeq HABDm3).
	assert(Hcomp : 3 <= 3) 
	       by (repeat constructor).
	assert(Hincl : incl (A:: B:: D:: nil) (A:: B:: C:: D:: nil)) 
	       by (repeat clear_all_rk;my_inO).
	apply (
	  rule_5 (A:: B:: D:: nil) (A:: B:: C:: D:: nil) 3 3 HABDmtmp Hcomp Hincl
	      ).
}
assert(HABCDM : rk(A:: B:: C:: D::nil) <= 3) 
       by (solve_hyps_max HABCDeq HABCDM3).
assert(HABCDm : rk(A:: B:: C:: D::nil) >= 1) 
       by (solve_hyps_min HABCDeq HABCDm1).
intuition.
Qed.

Lemma LABC : forall A B C D ,
rk(A:: C::nil) = 2 -> rk(A:: B:: D::nil) = 3 -> 
rk(C:: D::nil) = 2 -> rk(A:: C:: D::nil) = 2 -> 
rk(A:: B:: C::nil) = 3.
Proof.
intros A B C D 
HACeq HABDeq HCDeq HACDeq .

assert(HABCm2 : rk(A:: B:: C:: nil) >= 2).
{
	assert(HACmtmp : rk(A:: C:: nil) >= 2) 
	      by (solve_hyps_min HACeq HACm2).
	assert(Hcomp : 2 <= 2) 
	     by (repeat constructor).
	assert(Hincl : incl (A:: C:: nil) (A:: B:: C:: nil)) 
	     by (repeat clear_all_rk;my_inO).
	apply (
	  rule_5 (A:: C:: nil) (A:: B:: C:: nil) 2 2 HACmtmp Hcomp Hincl
	      ).
}
assert(HABCm3 : rk(A:: B:: C:: nil) >= 3).
{
	assert(HACDMtmp : rk(A:: C:: D:: nil) <= 2) 
	       by (solve_hyps_max HACDeq HACDM2).
	assert(HABCDeq : rk(A:: B:: C:: D:: nil) = 3) 
	       by 
	       (apply LABCD with (A := A) (B := B) (C := C) (D := D) ; assumption).
	assert(HABCDmtmp : rk(A:: B:: C:: D:: nil) >= 3) 
	       by (solve_hyps_min HABCDeq HABCDm3).
	assert(HACmtmp : rk(A:: C:: nil) >= 2) 
	       by (solve_hyps_min HACeq HACm2).
	assert(  Hincl : 
	         incl (A:: C:: nil) 
	         (list_inter (A:: B:: C:: nil) (A:: C:: D:: nil))) 
	       by (repeat clear_all_rk;my_inO).
	assert(  HT1 : 
	         equivlist (A:: B:: C:: D:: nil) 
	         (A:: B:: C:: A:: C:: D:: nil)) 
	       by (clear_all_rk;my_inO).
	assert( HT2 : 
	        equivlist (A:: B:: C:: A:: C:: D:: nil) 
	        ((A:: B:: C:: nil) ++ (A:: C:: D:: nil))
	      ) by (clear_all_rk;my_inO).
	rewrite HT1 in HABCDmtmp;rewrite HT2 in HABCDmtmp.
	apply (
	  rule_2 
	     (A:: B:: C:: nil) (A:: C:: D:: nil) (A:: C:: nil) 
	     3 2 2 HABCDmtmp HACmtmp HACDMtmp Hincl
	     ).
}
assert(HABCM : rk(A:: B:: C::nil) <= 3)
       by (solve_hyps_max HABCeq HABCM3).
assert(HABCm : rk(A:: B:: C::nil) >= 1) 
       by (solve_hyps_min HABCeq HABCm1).
intuition.
Qed.
\end{lstlisting}

\end{document}